\documentstyle[preprint,aps,epsfig]{revtex}
\tightenlines
\begin{document}
\draft

\title{Nuclear enhanced power corrections to DIS structure functions}
\author{Xiaofeng Guo$^{1}$\footnote{Current email address:
gxf@iastate.edu.}, Jianwei Qiu$^{2}$, and Wei Zhu$^{3,4}$} 
\address{$^1$Department of Physics and Astronomy,
             University of Kentucky,\\
             Lexington, Kentucky 40506, USA\\
         $^2$Department of Physics and Astronomy,
             Iowa State University \\
             Ames, Iowa 50011, USA \\
         $^3$Department of Physics,
             East China Normal University\\
             Shanghai 200062, China \\
         $^4$International Institute of Theoretical and Applied
             Physics, Iowa State University\\
             Ames, Iowa 50011, USA}
\date{October 1, 2001}
\maketitle
\begin{abstract}
We calculate nuclear enhanced power corrections to structure functions
measured in deeply inelastic lepton-nucleus scattering in Quantum
Chromodynamics (QCD).  We find that the nuclear medium enhanced power
corrections at order of $O(\alpha_s/Q^2)$ enhance the longitudinal
structure function $F_L$, and suppress the transverse structure
function $F_1$.  We demonstrate that strong nuclear effects in
$\sigma_A/\sigma_D$ and $R_A/R_D$, recently observed by HERMES
Collaboration, can be explained in terms of the nuclear enhanced
power corrections. 
\end{abstract}
\vspace{0.2in}



Recently, a strong nuclear dependence in lepton-nucleus deeply
inelastic scattering (DIS) was observed by HERMES collaboration
\cite{Hermes}. HERMES data show a large nuclear enhancement in the
longitudinal cross section when Bjorken $x_B$ is small.  The ratio of 
longitudinal cross sections $\sigma_L^{A}/\sigma_L^D$ measured by
HERMES Collaboration was as large as $2$ for scattering off a $^{14}$N
target. On the other hand, due to nuclear shadowing in the small $x$
region, the ratio of the transverse cross section
$\sigma_T^A/\sigma_T^D$ is much less than 1, which leads to a very
large nuclear enhancement in a double ratio 
$(\sigma_L^A/\sigma_T^A)/(\sigma_L^D/\sigma_T^D)$ \cite{Hermes}.  The
anomalous nuclear effects observed by HERMES have generated lots 
of interests \cite{theory}.  In this Letter, we calculate the nuclear 
dependence of the DIS cross sections (or DIS structure functions) at
low $Q^2$ in terms of medium enhanced power corrections in QCD
perturbation theory, and provide our understanding of HERMES data. 

For low energy fixed target DIS processes, like HERMES experiment, we
neglect a small contribution 
from vector boson $Z$ and take the one-photon exchange approximation.
Like other DIS experiments, HERMES experiment measures inclusive
lepton-nucleus cross sections as a function of $x_B$ and
virtual photon's invariant mass square $Q^2$ \cite{Hermes} 
\begin{equation} 
\frac{d\sigma_{\ell h}}{dx_B dQ^2}
=\frac{\sigma_{\rm Mott}}{E E'}\,
 \frac{\pi F_2(x_B,Q^2)}{\epsilon\; x_B} 
\left[
 \frac{1+\epsilon\; R(x_B,Q^2)}{1+R(x_B,Q^2)}
\right] ,
\label{DIS-cross-x}
\end{equation}
where $\sigma_{\rm Mott}$ represents an elastic cross section for
lepton scattering from a point charge, and $E$ and $E'$ are incoming
and scattered lepton energy, respectively.  The parameter $\epsilon$
is given by \cite{Hermes}
\begin{equation}
\epsilon = \frac{4(1-y)-Q^2/E^2}{4(1-y)+2y^2 + Q^2/E^2} \, ,
\label{epsilon}
\end{equation}
where $y=(E-E')/E$.  The $R(x_B,Q^2)$ in Eq.~(\ref{DIS-cross-x}) is
the ratio of longitudinal to transverse DIS cross sections
$R=\sigma_L/\sigma_T$.  When $x_B<0.03$, HERMES data on the ratio of
inclusive DIS cross sections from nucleus $A$ and deuterium
$D$(=$^2$H), $\sigma_A/\sigma_D$, show major differences from what
have been measured by other experiments in terms of the dependences on
$x_B$, $Q^2$, and $\epsilon$ \cite{Hermes}.  
When $x_B$ decreases, the ratio falls steeply, and is much
smaller than what was measured by other experiments. 
When $Q^2$ increases, the ratio is not only small but also decreases.
On the other hand, the ratio increases when $\epsilon$ increases. 
Why HERMES data is so different from other data in the small 
$x_B$ region?  

From Eq.~(\ref{DIS-cross-x}), the ratio of inclusive DIS cross
sections from nucleus $A$ and deuterium $D$(=$^2$H) can be expressed
as \cite{Hermes} 
\begin{equation}
\frac{\sigma_A}{\sigma_D}
= \frac{F_2^A}{F_2^D}\, 
\frac{(1+\epsilon\, R_A)(1+R_D)}
     {(1+R_A)(1+\epsilon\, R_D)} \, ,
\label{ratio-sigma}
\end{equation}
where $R_A$ and $R_D$ represent the ratio of $\sigma_L/\sigma_T$ for
nucleus $A$ and deuterium $D$.  HERMES Collaboration shows that the
ratio of $F_2^A/F_2^D$ extracted from its data is consistent with the
previous measurements, and attributes the differences in the ratio of
cross sections, $\sigma_A/\sigma_D$, to the strong nuclear dependence
in $R_A/R_D$ \cite{Hermes}.  At the same $x_B$, the key difference
between HERMES data and  other data is the range of $Q^2$ and
$\epsilon$.  When $x_B$ is small, the $Q^2$ of HERMES data is much
smaller than that of other data sets.  It is then natural to
investigate the role of $1/Q^2$ type power corrections to the ratio of
the DIS cross sections, $R$.
  
\begin{figure}
\begin{center}
\epsfig{figure=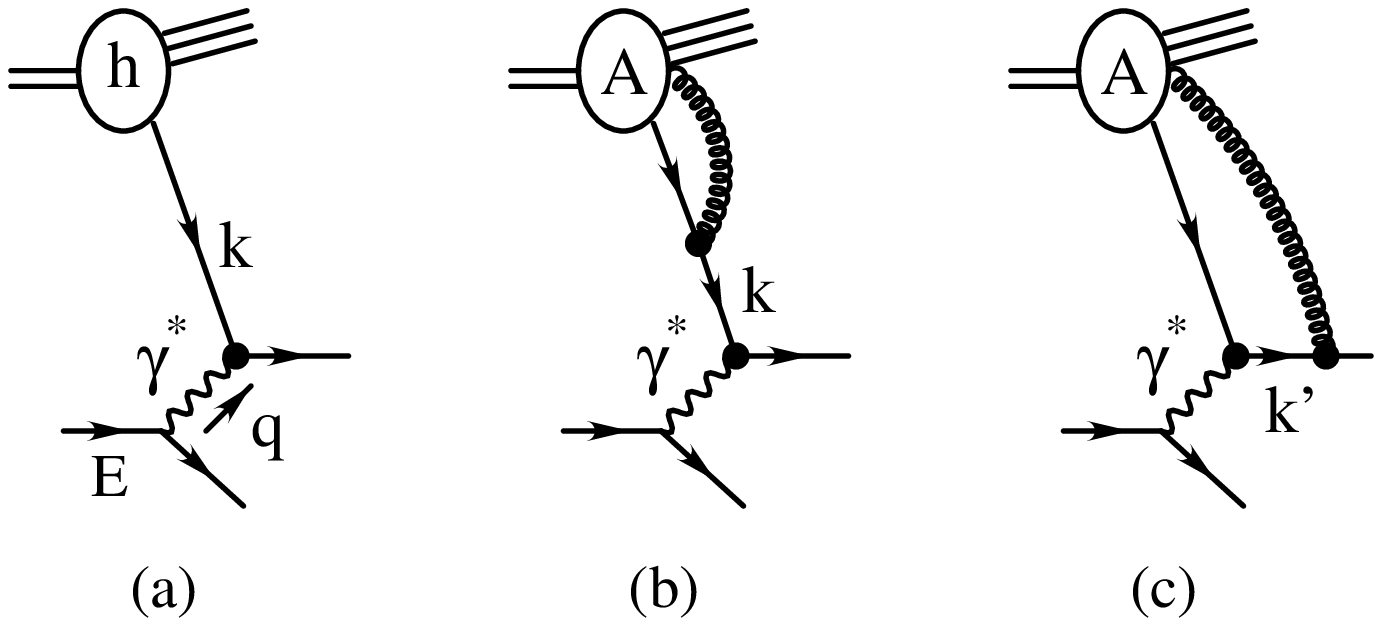,width=4.0in}
\end{center}
\caption{Sketch for lepton-hadron and lepton-nucleus DIS: (a)
single-parton scattering, (b) multiparton interaction internal to the
nucleus, and (c) multiparton interaction at the hard collision.}  
\label{fig1}
\end{figure}

When energy exchange of the collision, $Q$, is much
larger than the typical momentum scale of the hadron wave function, the
DIS process is dominated by a single hard collision between the
lepton and a quark via the virtual photon, as 
shown in Fig.~\ref{fig1}(a).  The corresponding cross section is given by
the probability to find a quark within a hadron times an elastic
cross section between this quark and the incoming lepton.  QCD radioactive
corrections to such a single parton scattering picture are
well-defined \cite{CTEQ-handbook}.  However, in a large
nucleus, partons from different nucleons can interact with each other
before the hard collision takes place, as shown in Fig.~\ref{fig1}(b),
or interact with the scattered quark after the hard collision, as
shown in Fig.~\ref{fig1}(c).  Just like the process in
Fig.~\ref{fig1}(a), leading contributions from the process shown
in Fig.~\ref{fig1}(b) is dominated by the phase space where the quark
of momentum $k$ is pinched to be on-mass shell.  That is, the
multiparton interactions from different nucleons are separated from 
the hard collision by a ``long'' time scale.  Therefore, such
multiparton interactions are internal to the nucleus, and do not
affect the short-distance hard collision between the quark and the
virtual photon.  On the other hand, such interactions are responsible
for the differences between the parton distribution in a nucleus and
that in a free nucleon \cite{MQ-recomb}.  Because of the leading twist
nature of the operators defining the parton distributions in a
nucleus, and the fact that $F_2$ is dominated by the contributions
from the parton distributions, the ratio of $F_2^A/F_2^D$ has a small
$Q^2$ dependence \cite{Qiu-shadowing,NMC-q2-dep}.   

Unlike the parton momentum $k$ in Fig.~\ref{fig1}(b), the parton
momentum $k'$ in Fig.~\ref{fig1}(c) is not pinched after we square the 
scattering amplitude due to different gluon momenta in the amplitude
and its complex conjugate \cite{QS-fac}.  The integration of the
unpinched parton momentum $k'$ can be deformed away from the $k'^2=0$,
and the quark-gluon interaction in Fig.~\ref{fig1}(c) is a part of the 
hard collision.  Because of the extra interactions with the scattered
quark, physical contributions from such multiparton scattering are
suppressed by inverse powers of $Q^2$, known as the high twist
contributions \cite{EFP-power,Qiu-power,QS-power}.  Although their
contributions to the DIS 
structure functions are normally small in lepton-hadron collisions,
such power corrections can be important in the 
nuclear collisions because of the enhancement of nuclear size
\cite{QS-hardprobe}.  In the rest of this Letter, we show that the
power corrections caused by the type of multiparton interactions 
shown in Fig.~\ref{fig1}(c) are responsible for the features seen 
in the HERMES data.

In terms of DIS structure functions, $F_1(x_B,Q^2)$ and
$F_2(x_B,Q^2)$, the ratios of the DIS cross sections can be expressed
as \cite{HM-book} 
\begin{equation}
\frac{\sigma_L}{\sigma_T}
= \frac{1}{2x_B F_1}\left[
  \left(1+\frac{4m_N^2 x_B^2}{Q^2}\right) F_2 - 2x_B F_1 \right] 
\equiv \frac{F_L}{2\, F_1}
\label{R-def}
\end{equation}
where $m_N$ is nucleon mass, and the longitudinal structure function
$F_L$ is defined as \footnote{Our definition of $F_L$ in
Eq.~(\ref{FL-def}) differs by a factor of $1/x_B$ from that 
in Ref.~\cite{Hermes}.}
\begin{equation}
F_L = \frac{1}{x_B} \left[
        \left(1+\frac{4m_N^2 x_B^2}{Q^2}\right) F_2 
        - 2x_B F_1 \right] 
    \approx \frac{1}{x_B} \left[F_2 - 2x_B F_1\right]
\label{FL-def}
\end{equation}
when $x_B\ll 1$ or $m_N^2\ll Q^2$.  Let us introduce two projection
operators,  
\begin{eqnarray}
e^{\mu\nu}_L 
&=& \frac{1}{Q^2}\left[q^\mu +2x_B p^\mu\right]
                  \left[q^\nu +2x_B p^\nu\right]\, ,
\label{el-def} \\
e^{\mu\nu}_T
&=& \frac{1}{p\cdot q}\left[p^\mu q^\nu + q^\mu p^\nu\right]
   +\frac{2x_B}{p\cdot q} p^\mu p^\nu - g^{\mu\nu}\, , 
\label{et-def}
\end{eqnarray}
where $q^\mu$ is virtual photon's four-momentum with $q^2=-Q^2$.
The DIS hadronic tensor $W^{\mu\nu}$ can be then expressed as
\cite{Qiu-power}  
\begin{equation}
W^{\mu\nu}(x_B,Q^2) = \frac{1}{2} \left[
                      e^{\mu\nu}_L\, F_L(x_B,Q^2)
                    + e^{\mu\nu}_T\, 2\, F_1(x_B,Q^2) \right]\, .
\label{W-munu}
\end{equation}
Therefore, we can extract the DIS structure functions from the
hadronic tensor as 
\begin{equation}
F_L(x_B,Q^2) = 2\, e^{\mu\nu}_L W_{\mu\nu}(x_B,Q^2), 
\quad \mbox{and} \quad
F_1(x_B,Q^2) = \frac{1}{2}\, e^{\mu\nu}_T W_{\mu\nu}(x_B,Q^2)\, .
\label{FL-FT-def}
\end{equation}
At the lowest order, the DIS structure functions in lepton-nucleus
collisions are given by
\begin{eqnarray}
F_L^A(x_B,Q^2) &=& 0\, ,
\label{FL-lo} \\
F_1^A(x_B,Q^2) &=& \frac{1}{2x_B}\, F_2^A(x_B,Q^2) 
     = \frac{1}{2}\, \sum_q e_q^2 \,
       \phi_{q/A}(x_B,Q^2)\, ,
\label{F1-lo} 
\end{eqnarray}
where $\sum_q$ runs over all quark and antiquark flavors and $e_q$ is
corresponding fractional charge.  The $\phi_{q/A}$ is an effective
quark distribution in a nucleus of atomic number $A$, which is
normalized by dividing the $A$. 

\begin{figure}
\begin{center}
\epsfig{figure=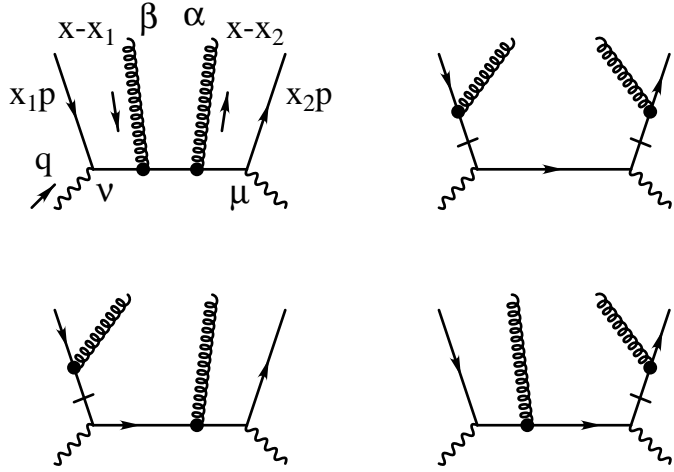,width=3.5in}
\end{center}
\caption{Leading order Feynman diagrams that contribute to 
the leading power corrections to the DIS structure functions.} 
\label{fig2}
\end{figure}

The lowest order power corrections to the DIS structure functions are
given by the Feynman diagrams in Fig.~\ref{fig2}, plus the diagrams
with four-quark lines instead of two-quark-two-gluon lines
\cite{Qiu-power}.  Since we are interested in the medium size enhanced
power corrections, we neglect the four-quark processes
\cite{QS-hardprobe}.  In Fig.~\ref{fig2}, the quark lines with a short
bar are special quark propagators, which are equal to the normal quark
propagators with the leading twist contributions removed
\cite{Qiu-power}.  The diagrams with special quark
propagators in Fig.~\ref{fig2} are important to preserve the
electromagnetic current conservation at the level of power
corrections.  Feynman rules for the special quark propagators
are given in Refs.~\cite{Qiu-power,BQ-dy-sts}.  In the
light-cone gauge of the strong interaction, contributions to the
hadronic tensor from the diagrams in Fig.~\ref{fig2} can be factorized
as \cite{QS-hardprobe}
\begin{equation}
W^{\mu\nu}_{qA}(x_B,Q^2) 
= \frac{1}{4\pi}\, \int dx\, dx_1\, dx_2\, T_{qA}(x,x_1,x_2) 
  H_{qA}^{\mu\nu}(x,x_1,x_2,q)\, .
\label{W-qA}
\end{equation}
The hadronic matrix element $T_{qA}$ is given by 
\begin{eqnarray}
T_{qA}(x,x_1,x_2)
&=&
\left(\frac{1}{x-x_1}\right) \left(\frac{1}{x-x_2}\right)
\int \frac{dy^-}{2\pi}\, \frac{dy_1^-}{2\pi}\,
      \frac{dy_2^-}{2\pi}\,
      {\rm e}^{ix_1p^+y^-}\, {\rm e}^{i(x-x_1)p^+y_1^-}\,
      {\rm e}^{-i(x-x_2)p^+y_2^-}
\nonumber \\
&\times &
\langle P_A | \bar{\psi}(0) \frac{\gamma^+}{2}
              F^{+\alpha}(y_2^-)F_{\alpha}^{\ +}(y_1^-)
              \psi(y^-) |P_A\rangle\, ,
\label{TqA}
\end{eqnarray}
where $p\equiv P_A/A$ is averaged momentum per nucleon.  
In deriving $T_{qA}$, we used $F^{+\alpha}(y^-)=n^\rho
\partial_\rho A^\alpha(y^-)$ in the light-cone gauge. 
The partonic part $H^{\mu\nu}_{qA}$ is given by the Feynman diagrams
in Fig.~\ref{fig2} with the quark lines contracted by
$\frac{1}{2}\gamma\cdot p$ and gluon lines contracted by $\frac{1}{2}
d_{\alpha\beta}$.  The transverse tensor
$d_{\alpha\beta}=-g_{\alpha\beta} + \bar{n}_\alpha n_\beta + n_\alpha
\bar{n}_\beta$ with two lightlike vectors, 
$\bar{n}^\mu=\delta^{\mu +}$ and $n^\mu=\delta^{\mu -}$. After
contracting with the projection operators, we obtain \cite{Qiu-power} 
\begin{eqnarray}
e^{\mu\nu}_L H_{\mu\nu}^{qA}(x_B,Q^2) 
&=& 4\, e_q^2 \left(\frac{4\pi^2\alpha_s}{3}\right)
    \frac{1}{Q^2}\, \delta(x-x_B)\, ,
\label{FL-hard} \\
e^{\mu\nu}_T H_{\mu\nu}^{qA}(x_B,Q^2) 
&=& 2\, e_q^2 \left(\frac{4\pi^2\alpha_s}{3}\right)
    \frac{1}{Q^2}\, x_B\left[
      \frac{\delta(x_2-x_B)}{x_1-x_B} 
     +\frac{\delta(x_1-x_B)}{x_2-x_B} \right]\, . 
\label{FT-hard}
\end{eqnarray}
By substituting the above partonic hard parts into Eq.~(\ref{W-qA}), we
can derive the leading power corrections to the DIS structure
functions defined in Eq.~(\ref{FL-FT-def}).

In order to derive the medium size enhanced power corrections, we need
to identify the position space integrals that can be extended freely
to the size of the target.  For the leading power corrections in
Eq.~(\ref{W-qA}), only position space integrals are from the
$dy_i^-$'s in Eq.(\ref{TqA}).  Generally, these $y_i^-$ integrals
cannot grow with the size of the target because of oscillation of the
exponential factors.  Because of the nature of Fourier transform, the
exponentials in Eq.~(\ref{TqA}) are linear functions of parton
momentum fractions.  Since the kinematics of inclusive DIS is
only sensitive to the total momentum from the target, two of
the three momentum fractions in Eq.~(\ref{W-qA}) cannot be fixed by
the hard collision.  As pointed out in Ref.~\cite{QS-hardprobe}, 
the integration of these two momentum fractions can be fixed by
two unpinched poles from the partonic parts, and the integration
removes the oscillation of two exponential factors.  Therefore, 
two corresponding $y_i^-$ integrations can be extended freely to the
whole target, and generate the target size enhanced power corrections.

After combining the hadronic matrix element in Eq.~(\ref{TqA}) with
the partonic hard parts in Eq.~(\ref{FL-hard}), the momentum fraction
$x$ is fixed by the $\delta(x-x_B)$, while $dx_1$ and $dx_2$
integrations are dominated by the region where $x_1\sim x_B$ and
$x_2\sim x_B$, because of the two poles at $1/(x_B-x_1)$ and
$1/(x_B-x_2)$ respectively.  The fact that two poles are from the matrix
element and do not have explicit $i\epsilon$ is an artifact of
choosing the light-cone gauge.  In order to recover the correct
$i\epsilon$ within the light-cone gauge calculations, we adapt the
$i\epsilon$ prescription introduced in Ref.~\cite{MQ-recomb},
\begin{equation}
\left(\frac{1}{x_B-x_1}\right) \left(\frac{1}{x_B-x_2}\right) 
\Longrightarrow
\left(\frac{1}{x_1-x_B+i\epsilon}\right) 
\left(\frac{1}{x_2-x_B-i\epsilon}\right) \, .
\label{i-epsilon}
\end{equation}
We can also work in a covariant gauge to avoid the spurious poles in
the light-cone gauge \cite{BQ-dy-sts,LQS-dis}.  After carrying out
$dx_1$ and $dx_2$ integrations, we obtain the leading power corrections
to the longitudinal structure function,
\begin{equation}
\left. F_L^A(x_B,Q^2)\right|_{1/Q^2} 
= \sum_q e_q^2\, 4 \left(\frac{4\pi^2\alpha_s}{3}\right)
    \frac{1}{Q^2}\, T_{qF}^A(x_B,Q^2)\, ,
\label{FL-power}
\end{equation}
where $\sum_q$ runs over all quark and antiquark flavors, and the
twist-4 quark-gluon correlation function is given by 
\cite{LQS-qg-dis} 
\begin{eqnarray}
T_{qF}^{A}(x_B,Q^2) 
&=&
 \int \frac{dy^{-}}{2\pi}\, e^{ix_Bp^{+}y^{-}}
 \int \frac{dy_1^{-}dy_{2}^{-}}{2\pi} \,
      \theta(y_1^{-}-y^{-})\,\theta(y_{2}^{-}) 
\nonumber \\
&\times &
     \langle p_{A}|\bar{\psi}_{q}(0)\,\frac{\gamma^{+}}{2}\,
                   F^{+\alpha}(y_{2}^{-})F_{\alpha}^{\ +}(y_{1}^{-})
                  \psi_{q}(y^{-})|p_{A} \rangle \ .
\label{TqF-A}
\end{eqnarray}
In order to derive leading power corrections to $F_1$, we expand the
$x_1$ and $x_2$ in Eq.~(\ref{FT-hard}) around $x$ because of the poles
in Eq.~(\ref{TqA}),
\begin{equation}
  \frac{\delta(x_2-x_B)}{x_1-x_B} 
 +\frac{\delta(x_1-x_B)}{x_2-x_B}
\approx 
 -\delta'(x-x_B) + ...
\label{approx-x}
\end{equation}
where ``...'' represents terms proportional to powers of $x_i-x$ with
$i=1,2$.  Since these terms cancel the poles in Eq.~(\ref{TqA}), we
can neglect them for calculating the leading nuclear size enhanced
contributions, and we obtain
\begin{equation}
\left. F_1^A(x_B,Q^2)\right|_{1/Q^2} 
= \frac{1}{2} \, \sum_q e_q^2
\left(\frac{4\pi^2\alpha_s}{3}\right)
    \frac{1}{Q^2}\, x_B\frac{d}{dx_B}T_{qF}^A(x_B,Q^2)\, ,
\label{FT-power}
\end{equation}
where $\sum_q$ runs over all quark and antiquark flavors, and
$T_{qF}^A$ is given in Eq.~(\ref{TqF-A}). 

In order to estimate the size and sign of the leading power
corrections to the structure functions in Eqs.~(\ref{FL-power}) and
(\ref{FT-power}), we need the knowledge of the twist-4 quark-gluon
correlation function $T_{qF}^A(x_B,Q^2)$.  Although this function is
nonperturbative, QCD factorization theorem enables us to find the
information on $T_{qF}^A$  from other physical observables.  For the
following numerical estimate, we adopt the model \cite{LQS-qg-dis} 
\begin{equation}
T_{qF}^A(x,Q^2) \approx \lambda^2\, A^{1/3}\, \phi_{q/A}(x,Q^2)\, ,
\label{TqF-model}
\end{equation}
and use $\lambda^2\approx 0.01$~GeV$^2$ extracted from the Drell-Yan
transverse momentum broadening in hadron-nucleus collisions
\cite{Guo-dy-qt}.  Substituting $T_{qF}^A$ in Eq.~(\ref{TqF-model})
into Eqs.~(\ref{FL-power}) and (\ref{FT-power}), we find that the
leading nuclear size enhanced power corrections to $F_L$ are positive, 
while the corrections to $F_1$ are negative.  Because of the opposite
signs, the nuclear enhanced power corrections to $F_2$ are smaller.  

In order to understand the HERMES data, we consider the following
double ratio,
\begin{equation}
R_R \equiv 
\left.
\frac{(1+\epsilon\,R_A)}
     {(1+R_A)}
\right/
\frac{(1+\epsilon\,R_D)}
     {(1+R_D)} \, .
\label{RR}
\end{equation}
Since $F_2^A/F_2^D$ measured by HERMES Collaboration is consistent
with the ratio measured by other experiments \cite{Hermes}, the double
ratio $R_R$ should be responsible for all unique features of HERMES
data, such as the $x_B$, $Q^2$, and $\epsilon$ dependence of the
ratio $\sigma_A/\sigma_D$, defined in Eq.~(\ref{ratio-sigma}). 
From Eq.~(\ref{R-def}), we have $R_A=F_L^A/(2F_1^A)$ with the power
corrections included for the DIS structure functions  
\begin{eqnarray}
F_i^A(x_B,Q^2) 
= \left. F_i^A(x_B,Q^2)\right|_{\rm LT}
+ \left. F_i^A(x_B,Q^2)\right|_{1/Q^2}
\label{Fi-def}
\end{eqnarray}
where $i=L,1$ and ``LT'' represents the leading twist contributions.
The leading power corrections to $F_L$ and $F_1$ are given in
Eqs.~(\ref{FL-power}) and (\ref{FT-power}), respectively.  The leading
order leading twist contributions are given in Eqs.~(\ref{F1-lo})
and (\ref{FL-lo}), respectively.  Since the leading order $F_L^A=0$
and the order of $\alpha_s$ contributions are renormalization scheme
independent, we use them as the nonvanishing contributions to $F_L$
\cite{ESW-book}  
\begin{eqnarray}
\left. F_L^A(x_B,Q^2)\right|_{\rm LT} 
&=& 
\frac{\alpha_s}{2\pi}\Bigg\{
\sum_q e_q^2 \int_{x_B}^1 \frac{dx'}{x'}\phi_{q/A}(x',Q^2) 
\left[\frac{8}{3}\left(\frac{x_B}{x'}\right)\right]
\nonumber \\
&+&
 \left(\sum_q e_q^2\right)
 \int_{x_B}^1 \frac{dx'}{x'}\phi_{g/A}(x',Q^2) 
 \left[2\left(\frac{x_B}{x'}\right)
        \left(1-\frac{x_B}{x'}\right)\right]
\Bigg\}.
\label{FL-lt}
\end{eqnarray}
Since $R_D$ is well measured and $R_D\approx R_P$ of proton
\cite{RD-param}, we let $R_D=F_L^D/(2F_1^D)$ with only the leading
twist contributions.  We find that our $R_D$ in low $Q^2$ and low
$x_B$ region are consistent with the parameterization in
Ref.~\cite{RD-param}.  Because we do not have well-tested parton
distributions at $Q^2$ as low as 0.5~GeV$^2$, we will use a set of
generic parton distributions in Ref.~\cite{QS-sts-photon} for the
following numerical estimate.

\begin{figure}
\begin{center}
\begin{minipage}[t]{3.0in}
\epsfig{figure=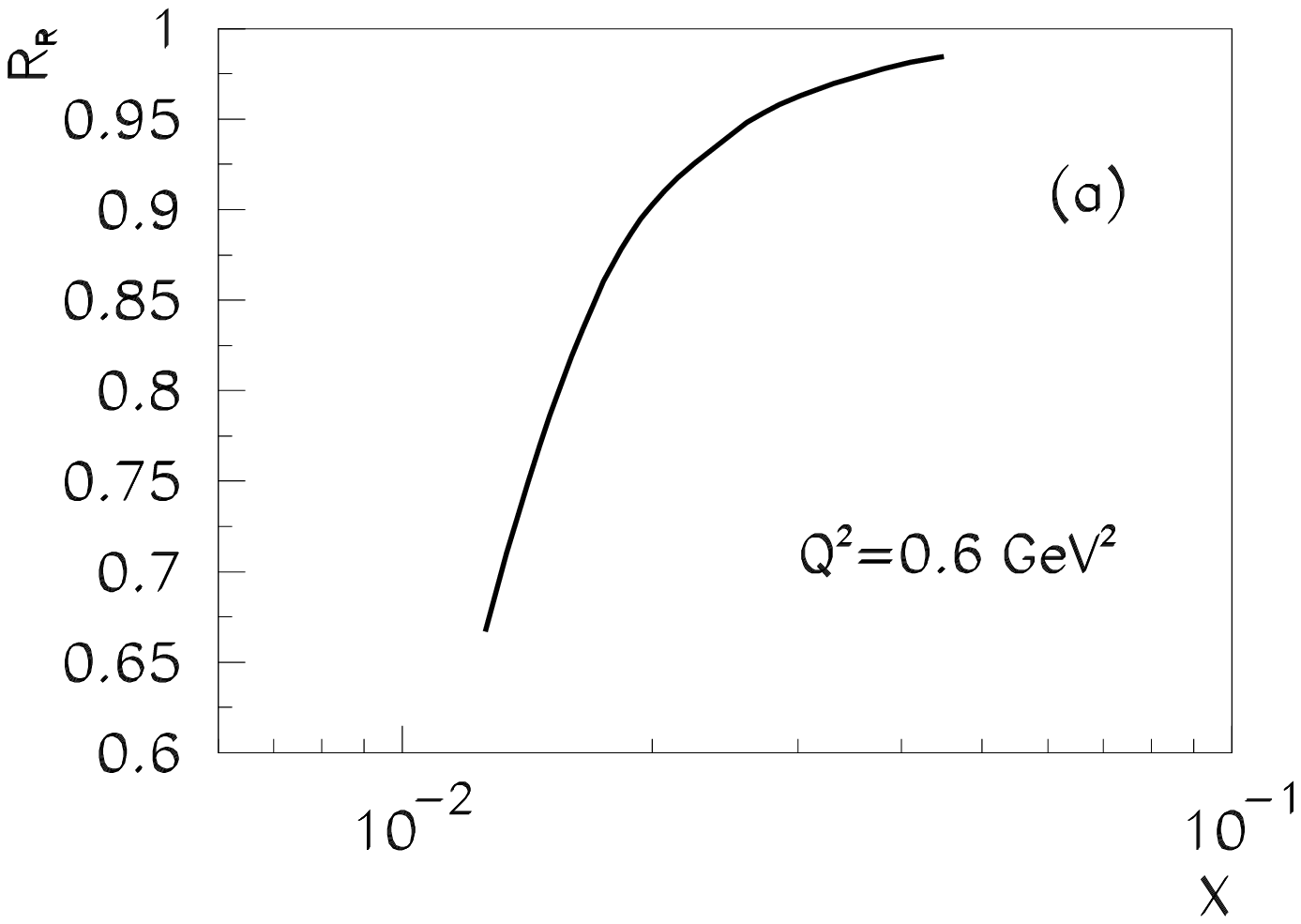,width=2.8in}
\end{minipage}
\hfill
\begin{minipage}[t]{3.0in}
\epsfig{figure=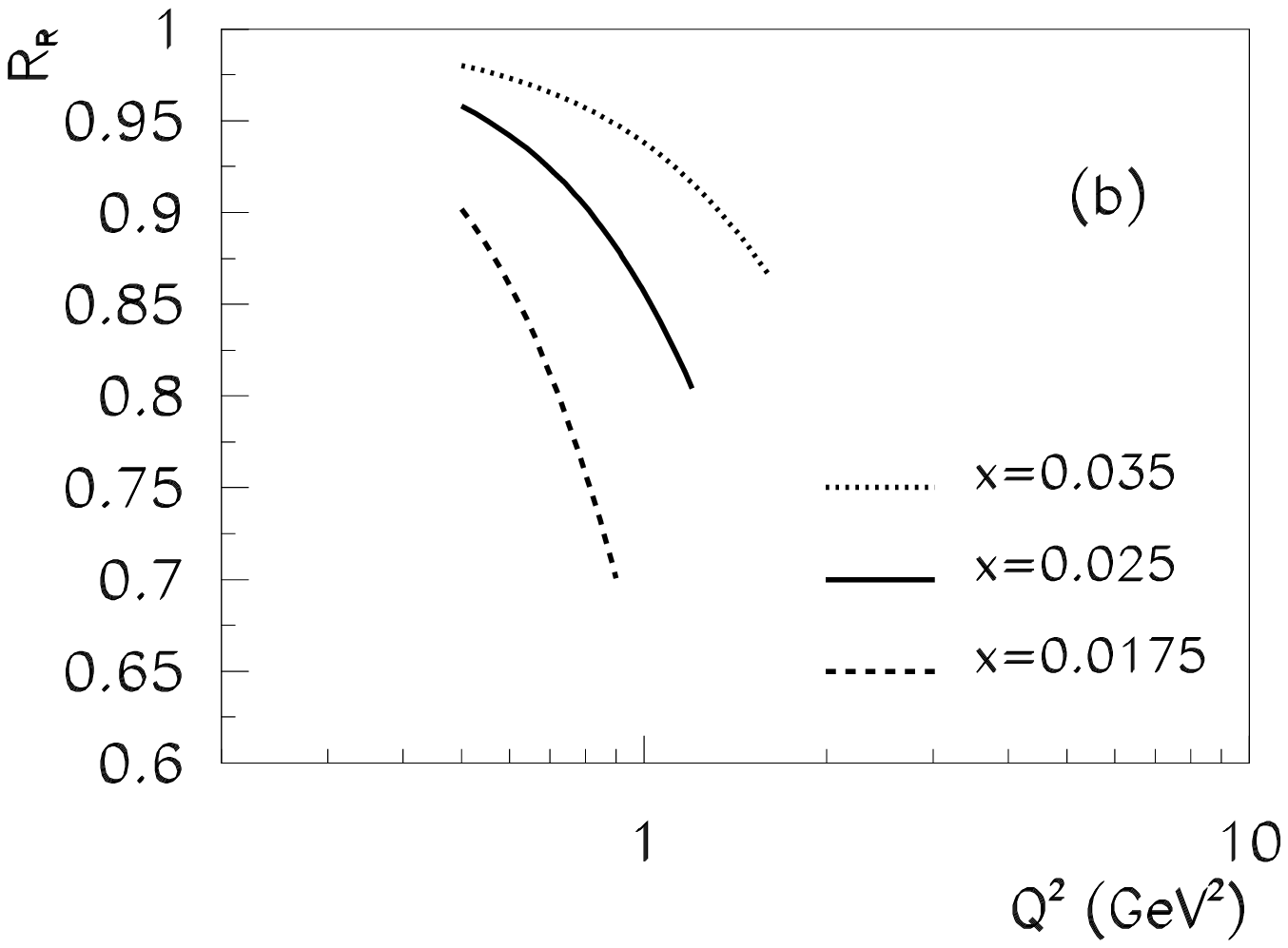,width=2.8in}
\end{minipage}
\end{center}
\caption{The double ratio $R_R$ in Eq.~(\protect{\ref{RR}}) as a
function of $x_B$ (a) and $Q^2$ (b).  The curves are explained in the
text.} 
\label{fig3}
\end{figure}

With the model of $T_{qF}^A$ in Eq.~(\ref{TqF-model}) and
$\lambda^2=0.01$~GeV$^2$, and without any further free parameter, we
plot the $x_B$ and $Q^2$ dependence of the double ratio $R_R$ for
$^{14}$N in Fig.~\ref{fig3}(a) and \ref{fig3}(b), respectively.  
In plotting Fig.~\ref{fig3}, we used $E=27.5$~GeV for the lepton beam
energy.  In Fig.~\ref{fig3}(a), the ratio is sensitive to the value of
$Q^2$, and $Q^2=0.6$~GeV$^2$ was used.  In Fig.~\ref{fig3}(b), three
curves correspond to $x_B = 0.0175$ (dotted), $0.025$ (solid), 
and $0.035$ (dashed), respectively. 
Clearly, Figs.~\ref{fig3}(a) and \ref{fig3}(b) reproduce all the key
features of the HERMES data shown in Fig.~1 and 2 of
Ref.~\cite{Hermes}.  The detailed numerical comparison with the HERMES
data requires better knowledge of original data analysis and reliable
low $Q^2$ parton distributions, and will be presented elsewhere.    

In the rest of the Letter, we explain why the nuclear enhanced power
corrections are consistent with the HERMES data.  As shown in
Fig.~\ref{fig3}(a) or Fig.~1 of Ref.~\cite{Hermes}, the ratio $R_R$ or
$\sigma_A/\sigma_D$ shows a steep falling feature when $x_B$ decreases, 
which is not observed in other experiments.  We find that the steep falling
was mainly caused by strong $x_B$ dependence of the $\epsilon$ and a
large $R_A$.  Substituting $y=Q^2/(2m_N x_B E)$ into
Eq.~(\ref{epsilon}), we find that $\epsilon$ falls steeply from 0.9 to
0.25 when $x_B$ decreases from 0.03 to 0.0125 at $Q^2=0.6$~GeV$^2$.
Thus, the ratio $(1+\epsilon\, R_A)/(1+R_A)$ falls along with the
$\epsilon$.   

The $Q^2$ dependence, as shown in Fig.~\ref{fig3}(b) or in Fig.~2 of
Ref.~\cite{Hermes}, is the most counterintuitive.  Normally, when
$Q^2$ increases, we expect the effect of power corrections to go away,
or the ratio $R_R$ to approach one.  However, the HERMES data shows
an opposite behavior.  We find that such a counterintuitive behavior
in HERMES data was caused by the strong $Q^2$ dependence of the
$\epsilon$ parameter.  For example, when $Q^2$ changes from
0.5~GeV$^2$ to 0.9~GeV$^2$, the $\epsilon$ at $x_B=0.0175$ decreases
from 0.78 to 0.13 by dropping a factor of 6, while the size of power 
corrections changes for less than a factor of 2.  We find that the
decrease of the double ratio $R_R$ when $Q^2$ increases is due to the
steep falling of the combination $\epsilon\, R_A$ in Eq.~(\ref{RR}). 

Similarly, the $\epsilon$ dependence of the HERMES data, shown in
Fig.~3 of Ref.~\cite{Hermes}, is a direct consequence of explicit
$\epsilon$ dependence in Eq.~(\ref{ratio-sigma}) or Eq.~(\ref{RR}).
From our numerical estimate, we also find that the observed large
double ratio $R_A/R_D \sim 5$ for $0.01< x < 0.03$, as shown in Fig.~5
of Ref.~\cite{Hermes}, is due to a strongly nuclear enhanced $R_A\sim
1.6$ at $Q^2=0.6$~GeV$^2$ and a relatively small $R_D\sim 0.32$, which
is consistent with other measurements \cite{RD-param}.

We conclude that strong nuclear effects in $\sigma_A/\sigma_D$ and
$R_A/R_D$, observed by HERMES Collaboration, are direct consequences
of strong nuclear enhanced power corrections. We demonstrated that our
numerical estimate reproduces all the features of the HERMES data.
When the $Q^2$ is as small as 0.6~GeV$^2$, we find that the structure
function $F_1(x_B,Q^2)$ is still dominated by the leading twist
contributions, while the longitudinal structure function
$F_L(x_B,Q^2)$ is dominated by the nuclear enhanced power corrections. 
Since the leading power corrections are large, it is therefore
important to study the nuclear enhanced power corrections at even
higher powers of $1/Q^2$ to have a better QCD prediction.


\section*{Acknowledgment}

One of us (W.Z.) thanks International Institute of Theoretical and
Applied Physics and nuclear theory group at Iowa State University for
support and hospitality when this work was initiated.
This work was supported in part by the U.S. Department of Energy under
Grant Nos. DE-FG02-87ER40731 and DE-FG02-96ER40989, and in part by the
National Science Foundation of China 10075020. 



\end{document}